\documentstyle[12pt,axodraw,epsfig,psfrag]{article}
\def\simlt{\stackrel{<}{{}_\sim}}
\def\simgt{\stackrel{>}{{}_\sim}}

\def\mkstop{m_{\tilde t, k}}

\newcommand\be{\begin{equation}}
\newcommand\ee{\end{equation}}
\newcommand\bea{\begin{eqnarray}}
\newcommand\eea{\end{eqnarray}}
\newcommand\ba{\begin{array}}
\newcommand\ea{\end{array}}

\begin{document}
\newcommand{\gsim}{ \mathop{}_{\textstyle \sim}^{\textstyle >} }
\newcommand{\lsim}{ \mathop{}_{\textstyle \sim}^{\textstyle <} }
\newcommand{\vev}[1]{ \left\langle {#1} \right\rangle }
\newcommand{\bra}[1]{ \langle {#1} | }
\newcommand{\ket}[1]{ | {#1} \rangle }
\newcommand{\EV}{ {\rm eV} }
\newcommand{\KEV}{ {\rm keV} }
\newcommand{\MEV}{ {\rm MeV} }
\newcommand{\GEV}{ {\rm GeV} }
\newcommand{\TEV}{ {\rm TeV} }


\begin{titlepage}

\begin{flushright}
hep-ph/0205010\\
SHEP 02-06
\end{flushright}

\vskip 0.4cm

\begin{center}

{\Large \bf  Supersymmetric Higgs Bosons in a 5D Orbifold Model}

\vskip 1.0cm

\def\thefootnote{\fnsymbol{footnote}}
{\bf 
V. Di Clemente, S. F. King and D. A. J. Rayner
}

\vskip 0.5cm

 {\it Department of Physics and Astronomy,
University of Southampton, \\
Southampton, SO17 1BJ, U.K.}\\

\vskip 1.0cm

\abstract{We analyze the phenomenology of the Higgs sector in a 5D model
compactified on an $S_1/Z_2$ orbifold where the compactification scale 
$M_C$ is around the TeV scale. We show that the conventional MSSM Higgs 
boson mass bounds in 4D can be violated when we allow the gauge sector, Higgs
and third family multiplets to live in the fifth extra dimension. 
Supersymmetry is broken at an orbifold fixed point which is spatially 
separated from the Yukawa brane where two chiral families are localized.
When the brane-localized supersymmetry breaking term for the stop sector is
arbitrarily large, we find that the stop KK-mode mass spectrum is completely
independent of the Higgs fields.  Hence, the Higgs
masses only receive radiative contributions from the top KK-modes.   
The 1-loop effective potential is insensitive to the cutoff scale of the 
theory and yields a negative Higgs squared-mass contribution that triggers 
electroweak symmetry breaking in the range $1.5\simlt\tan\beta\simlt 20$, 
where bottom/sbottom loop effects can be ignored. 
The recent LEP Higgs bound at $m_{h^{0}} > 114.1$ GeV, 
in conjunction with naturalness arguments, allows us to bracket the 
compactification scale $1.5 \simlt M_{C} \simlt 4$ TeV.  Within this 
parameter space, we find that the lightest Higgs boson mass has an upper 
bound $m_{h^{0}} \simlt 160$ GeV with the magnitude of the 
$\mu$-parameter restricted to the range $33 \simlt \left| \mu \right|
\simlt 347$ GeV.}

\end{center}
\end{titlepage}

\section{Introduction}

Extra-dimensional supersymmetric models with a TeV compactification 
scale~\cite{tev} have recently offered an exciting new environment for 
investigating electroweak symmetry breaking 
(EWSB) \cite{delgado,barbieri,arkani,quiros,dkr}. 
The main features of these types of models are the following: (i) they
provide a new mechanism to break supersymmetry (SUSY); (ii) the contribution
of quark/squark Kaluza-Klein (KK) modes to the Higgs mass term is 
negative, thus triggering EWSB at the TeV scale; (iii) the
1-loop radiative correction to the effective potential is free of
ultraviolet divergences, which implies that the Higgs physics is completely 
independent of the high-energy physics above some cutoff scale.
Remarkably, the requirement of 
${\mathcal N}=1$ SUSY in 5D (which is equivalent to ${\mathcal N}=2$ SUSY in 
4D after compactification ) 
leads to a finite 1-loop effective potential because even though SUSY
is globally softly broken, it is still preserved 
locally~\cite{arkani,masiero}~\footnote{Recently, the finiteness of this
kind of theory has been showed explicitly at 2-loops~\cite{von}.}.

Various models have been proposed to study EWSB in extra dimensions and their
phenomenological
implications~\cite{delgado,barbieri,arkani,quiros,dkr}. However 
we will concentrate only on those models which recover the MSSM below the 
compactification scale and are anomaly-free after the 
orbifold compactification~\cite{anomalous}.  The authors of 
Ref.~\cite{arkani} have made a qualitative study of the Higgs physics in 
their model, but there is little discussion of the
phenomenological implications. In contrast, the authors
of Ref.~\cite{quiros} have made an extensive study of 
Higgs phenomenology in their model and consider 
how radiative corrections modify the quadratic and 
quartic couplings of the tree-level potential.  However, their
approach neglects the non-renormalizable operators that arise at higher-order 
in the expansion of the full effective potential at 
1-loop and, as discussed in the Appendices, this brings into 
question the reliability of their calculation.
    
In this paper we show that the conventional MSSM lightest Higgs
boson mass bound $m_{h^{0}} \simlt 135$ GeV ~\cite{mssm} can be
violated by allowing some of the fields to live
in a fifth extra dimension~\footnote{Higher upper bounds on the
lightest boson mass have been recently calculated in the context of
SUSY in warped extra dimensions~\cite{casas} and a (de)constructed
model~\cite{grojean}.}.  Motivated by fine-tuning arguments, we are
led to an upper bound on the 
compactification scale $M_C \simlt 4$ TeV.  At this compactification scale,
the lightest Higgs boson mass can be pushed as high as 
$m_{h^{0}} \sim 160 \, {\mathrm GeV}$. 
We find that 
in order to have EWSB through radiative corrections, $\tan\beta$ can have a 
wide range $1.5 \simlt \tan\beta \simlt 20$ rather than the small 
range allowed by Ref.~\cite{quiros} $35 \simlt \tan\beta \simlt 40$.  We only
limit $\tan\beta \simlt 20$ since we neglect bottom-sbottom loop
effects.  Note that, unlike the MSSM, $\tan\beta \approx 1.5$ is not
ruled out by experiment in our extra dimensional model.

The layout of the remainder of the paper is as follows.  In section 
\ref{sec:setup} we review our 5D orbifold model, and in section 
\ref{sec:kkspec} we discuss the top/stop sector KK mass spectra.  
In section \ref{sec:higgs} we discuss the Higgs SUSY breaking parameters.
In section \ref{sec:pert} we discuss
how perturbativity and naturalness provide a constraint on the
theory cutoff $M_{\ast}$.  Then in section \ref{sec:2hdm}
we minimize the 1-loop effective potential and calculate the 
Higgs mass eigenvalues.  We find that experimental data and fine-tuning 
arguments allow us to constrain the physical parameter space.  Section 
\ref{sec:conc} concludes the paper.

\section{Our Model}  \label{sec:setup}

In this section we review some basic features of our model~\cite{dkr} 
in which the extra dimension of a 5D theory is compactified on an 
$S^{1}/Z_{2}$ orbifold. 
\begin{figure}[h]   
 \begin{center}  
  \begin{picture}(420,190)(-25,0)
   \Line( 100, 110 )( 145, 185 )
   \Line( 100,  15 )( 145,  90 )
   \Line( 100,  15 )( 100, 110 )
   \Line( 145,  90 )( 145, 185 )
   \Text( 100,   5 )[c]{$y=0$}
   \Text(  90, 110 )[r]{``Yukawa brane''}
   \Text(  90,  90 )[r]{1st/2nd family - 
                         $\hat{F}_{1,2L} \, , \, \hat{F}_{1,2R}$}
   \Text(  90,  70 )[r]{MSSM matter fields}
   \Text(210,160)[]{{\large $\hat{\mathcal V}_{MSSM}$}}
   \Text(200,130)[]{{\large $\hat{F}_{3L} \hspace*{1cm} \hat{F}_{3R}$}}
   \Text(200,100)[]{{\large $\hat{F}_{3L}^{m c} \hspace*{1cm} 
                              \hat{F}_{3R}^{m c}$}}
   \Text(195,70)[]{{\large $\hat{H}_{u} \hspace*{1.2cm} \hat{H}_{d}$}}
   \Text(200,40)[]{{\large $\hat{H}_{u}^{m c} \hspace*{1cm}
                              \hat{H}_{d}^{m c}$}}
   \Line( 255, 110 )( 300, 185 ) 
   \Line( 255, 15 )( 300, 90 )
   \Line( 255, 15 )( 255, 110 ) 
   \Line( 300, 90 )( 300, 185 )
   \Text( 255, 5 )[c]{$y= \pi R$} 
   \Text( 310, 110 )[l]{``source brane''} 
   \Text( 310, 90 )[l]{SUSY breaking sector} 
   \Text( 310, 70 )[l]{gauge singlet $\hat{S}$} 
  \end{picture}
 \end{center} 
  \caption{{\small Our model showing the parallel 3-branes spatially 
separated along the extra dimension y.  This extra dimension is compactified
on the orbifold $S^{1}/Z_{2}$ that leads to two fixed points at $y=0, \pi R$,
where the two 3-branes are located.  The first two chiral families 
($\hat{F}_{1,2}$) live on 
the Yukawa brane at $y=0$, while SUSY is broken by the F-term of a gauge 
singlet field ($\hat{S}$) on the source brane at $y=\pi R$.  
The third family ($\hat{F}_{3}$), gauge sector ($\hat{\mathcal V}_{MSSM}$) and
Higgs superfields ($\hat{H}_{u,d}$) live in the extra dimensional bulk along 
with their ${\mathcal N}=2$ SUSY partners which are required for 
consistency.}}
  \label{fig:setup}
\end{figure}
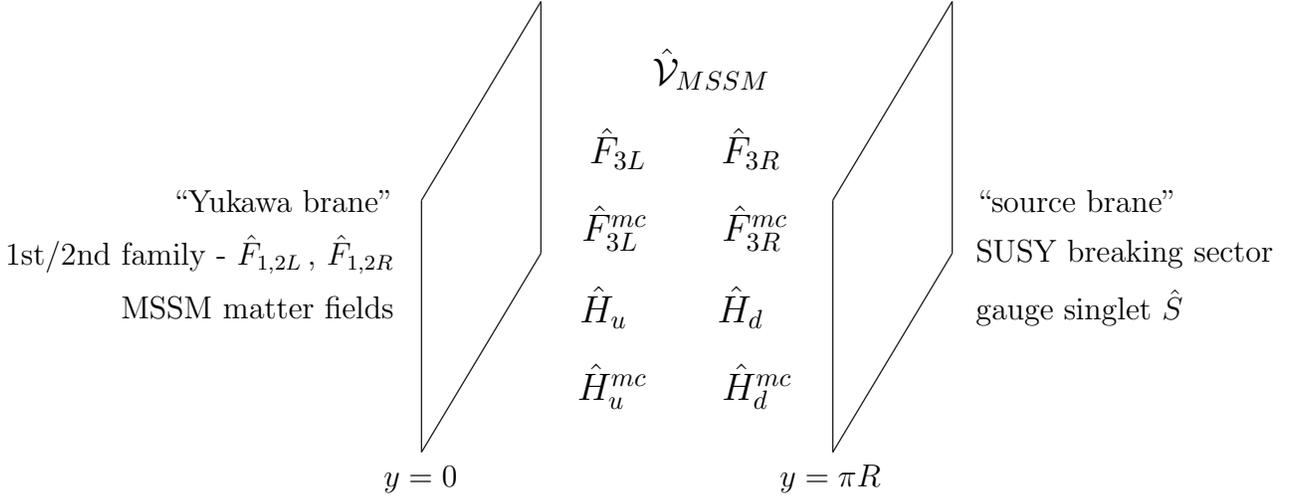

The orbifold compactification leads to a description
of 5D bulk fields as infinite
towers of 4D KK-modes, and a classification of bulk fields into odd
and even $Z_{2}$ parity~\footnote{Note that only even parity bulk fields 
have $k=0$ KK-modes and have a non-vanishing wavefunction profile at the fixed
points.}.  The $k^{th}$ KK-mode has a mass 
$m_{k} \sim {\mathcal O} \left( k \, M_{C} \right)$, where $M_C = 1/R$ is
the compactification scale, and the lowest KK resonance (k=0) can be
identified with the usual MSSM fields.  It is conventional in five-dimensional
orbifold models to exploit the equivalence of ${\mathcal N}=1$ SUSY
in 5D and ${\mathcal N}=2$ SUSY in 4D~\cite{peskin}.  Therefore, an 
${\mathcal N}=1$
multiplet in 5D can be decomposed into a 4D ${\mathcal N}=1$ chiral
multiplet and its CP-conjugate ``mirror'' which together form a
complete ${\mathcal N}=2$ hypermultiplet in 4D.
Similarly, an ${\mathcal N}=1$ vector multiplet in 5D can  be
decomposed into a vector and chiral multiplet in 4D~\footnote{The MSSM
vector multiplets $\hat{\mathcal V}_{MSSM}$ each contain a 4D
gauge field ($A_{\mu}$), two symplectic Majorana gauginos
($\lambda_{1,2}$) and a real scalar ($\Sigma$) ~\cite{peskin}.}.
The setup is given in Figure \ref{fig:setup} and 
shows the location of the superfields within the extra-dimensional space.

We localized 4D 3-branes at 
the two fixed points $y=0,\pi R$ arising from the orbifold
compactification.  SUSY is broken on the ``source''
brane at $y=\pi R$ when a localized gauge field singlet ($\hat{S}$)
acquires a non-zero F-term vacuum expectation value (VEV) $F_S$.  
The first two MSSM ($\hat{F}_{1,2}$) families are confined to the ``Yukawa'' 
brane~\footnote{This is known as the Yukawa brane since the superpotential
cannot be defined in the ${\cal N} =2$ supersymmetric bulk, and we are forced
to localize the ${\cal N} =1$ supersymmetric Yukawa couplings on the 4D brane
at $y=0$.} at $y=0$, 
while the third family ($\hat{F}_{3}$), MSSM gauge sector 
($\hat{\mathcal V}_{MSSM}$) and Higgs superfields ($\hat{H}_{u,d}$) live in 
the extra dimensional bulk~\footnote{The presence of
the top/stop fields in the bulk is phenomenologically important for their 
dominant 1-loop contribution to the Higgs potential.}.  
These bulk fields acquire tree-level soft
parameters due to their direct coupling with the SUSY breaking brane through
non-renormalizable operators defined at $y=\pi R$:
\begin{eqnarray}
 \delta {\cal L}_{\pi R} = -\delta(y-\pi R) \int \!\! d^4\theta \left[
  \frac{1}{M_{\ast}^{3}} \left[ c_{F_{3}} \hat{F}_{3}^{\dagger} \hat{F}_{3}
   +  c_{H_{u}} \hat{H}_{u}^{\dagger} \hat{H}_{u} 
    + c_{H_{d}} \hat{H}_{d}^{\dagger} \hat{H}_{d}
  + \left( c_{B\mu} \hat{H}_{u} \hat{H}_{d} + h.c. \right) \right] 
   \hat{S}^{\dagger} \hat{S}  \right. \nonumber \\
 + \left. \left( \frac{c_{\mu}}{ M_{\ast}^{2}} \hat{H}_{u} \hat{H}_{d} 
  \hat{S}^{\dagger} + h.c. \right) \right]  \hspace*{1.5cm}  \label{nonren}
\end{eqnarray}
where $M_*$ is the cutoff of the theory, and is often identified with 
the string scale. The coefficients 
$c_{F_{3}} (c_{H_{u,d}} , c_{\mu} , c_{B\mu})$ are the couplings of the third 
family (Higgs) superfields to the SUSY breaking sector.  There are also 
higher-dimensional Yukawa couplings localized on the Yukawa brane at $y=0$, 
but we will only consider the dominant top/stop sector couplings:
\begin{eqnarray}
 \delta {\cal L}_{0}^{Yuk} = -\delta(y) \, \frac{f_{t}}{M_{\ast}^{3/2}}
  \int \!\! d^{2}\theta \left( \hat{Q}_{3 L} \, \hat{H}_{u} \, 
   \hat{U}_{3R}^{c} + h.c. \right)   \label{yukawa}
\end{eqnarray}
where $f_{t} = (\pi R M_{\ast})^{3/2} \, y_{t}$ and $f_{t} (y_{t})$ is the 
5d (4d) Yukawa coupling.  The zero modes of
the neutral components of the two complex (scalar) Higgs doublets can be
expanded in terms of real and imaginary parts
\be
 H_{u}^{0}=\frac{1}{\sqrt{2}} \left( h_{u}+i \chi_{u} \right) \quad\quad
 H_{d}^{0}=\frac{1}{\sqrt{2}} \left( h_{d}+i \chi_{d} \right) 
\label{eq:higgsdef}
\ee
and electroweak symmetry is spontaneously broken when the real parts acquire
non-zero VEVs $<h_u>,<h_d> \neq 0$ and $<\chi_u> = <\chi_d> = 0$. 

This setup is like the gaugino mediated SUSY breaking ($\tilde{g}$MSB) 
scenario~\cite{gaugino}, but with the third family superfields moved 
from the Yukawa brane and placed in the extra-dimensional bulk.  The spatial 
separation of MSSM fields on the Yukawa brane away from the SUSY breaking 
sector alleviates the supersymmetric flavour-changing neutral-current (FCNC) 
problem since squark masses, arising from direct couplings between the two 
sectors, are exponentially suppressed by the separation between branes.  
Instead, first and second family squark masses are generated via flavour-blind
loop-effects~\footnote{Notice that the third family squarks acquire 
unsuppressed masses due to their direct coupling to the SUSY breaking.  
However, this does not undermine the solution to the FCNC problem since there
are much weaker experimental constraints on third family 
contributions~\cite{fcnc}.}.
Unlike previous models, our setup is motivated by a type I 
string-inspired model~\cite{rayner} in which the separation of the 
third family is generic, and often leads to realistic Yukawa textures.  

\subsection{Top/Stop KK Mass Spectra}  \label{sec:kkspec}

We will summarize the KK mass spectra found in our earlier paper~\cite{dkr}.  
The non-renormalizable operators in Eq.(\ref{nonren}) contain a delta-function
that induces mixing between different KK-modes, where the mixing strengths 
($\alpha_{\tilde{t}} , \alpha_{H}$) are proportional to the SUSY breaking 
VEVs:
\begin{equation}
 \alpha_{\tilde t} = c_{\tilde t}\pi\left(\frac{F_{S,\tilde{t}}^2}{M_*^4}
  \right) \left(\frac{M_*}{M_C}\right)  \hspace*{1cm} , \hspace*{1cm} 
 \alpha_{H} = c_{H}\pi \left(\frac{F_{S,H}^{2}}{M_{\ast}^{4}}\right) 
  \left(\frac{M_{\ast}}{M_C} \right)    \label{alphas}
\end{equation}
where $F_{S,\tilde{t}} (F_{S,H})$ is the F-term VEV associated with the 
singlet $\hat{S}$ field that couples to the stop (Higgs) 
fields~\footnote{The extra index on the SUSY breaking VEVs 
allows for non-universal couplings in the hidden sector 
(${\mbox F_{S,H} \neq F_{S,\tilde{t}}}$).}.  We have made 
the simplifying assumption that $c_{H_{u}} = c_{H_{d}} = c_{H}$.  The 
non-trivial mixing between different KK-modes requires that we diagonalize an 
infinite mass matrix to obtain the KK mass eigenvalues~\footnote{Details of
the diagonalization procedure are given in our earlier
paper~\cite{dkr}.}.  However, if the mixing is small, the mass matrix is
dominated by the diagonal components. In order to respect ${\mathcal N}=1$ 
SUSY transformations on the brane, we use an off-shell formulation of 
${\mathcal N}=2$ SUSY
in the 5D bulk~\cite{peskin} that mixes fields of different
$Z_{2}$-parity and so the KK-summation runs over the full tower 
($-\infty<k<\infty$).

The field-dependent top KK-mode mass eigenvalues are given by~\footnote{Notice
that the argument of the arctan function differs by a factor of $2\sqrt{2}$ 
compared to the previously published result~\cite{dkr}.}:
\be
 m_{t,k}[h_{u}] = \left| k M_{C} + \frac{M_{C}}{\pi}
   \arctan\left(\frac{y_{t} \, h_{u} \, \pi}{\sqrt{2} M_{C}}\right) \right| 
 \quad\quad  (k= -\infty, \dots, \infty)
\label{eq:topmass}
\ee
where the eigenvalues only depend on the real part of $H_{u}^{0}$.  The 
observable top mass is identified with the $k=0$ KK-mode,
\begin{eqnarray}
 m_{t}[h_u] = \frac{M_{C}}{\pi} \arctan \left(\frac{y_{t} \, h_{u} \, 
  \pi}{\sqrt{2} M_{C}} \right)   \label{eq:phystop}
\end{eqnarray} 
Notice that we can recover the usual MSSM relation 
$m_t[h_u] = y_t h_u/\sqrt{2}$ in the limit that ${\mbox M_C \rightarrow
\infty}$.  Eq.(\ref{eq:phystop}) is different due to the 
non-trivial mixing between KK-modes on the Yukawa brane.

The field-dependent stop KK mass eigenvalues $\mkstop[h_{u}]$ are solutions 
of the following transcendental equation
\be
 \frac{\pi \, \mkstop[h_{u}]}{M_C} \, \left[  
  \tan \left( \frac{\pi \, \mkstop[h_{u}]}{M_C} \right) 
   - \left( \frac{\pi \, y_t \, h_u}{\sqrt{2} M_C} \right)^{2} 
    \, \cot \left( \frac{\pi \, \mkstop[h_{u}]}{M_C} \right) \right] =
 \alpha_{\tilde t} \left[  
  1 +  \left( \frac{\pi \, y_t \, h_u}{\sqrt{2} M_C} \right)^{2} \right]
\label{x}
\ee
which can be solved by considering different limits of mixing 
$\alpha_{\tilde{t}}$. 

Minimal mixing ($\alpha_{\tilde t} \ll 1$) is equivalent to a very small
extra dimension ($M_{C} \sim 10^{16}$ GeV) where we expect to recover the
MSSM results.  The stop KK mass
eigenvalues are~\footnote{This case is particularly
interesting for grand unified theories (GUT) in extra
dimensions~\cite{hall} since it is possible to generate soft
parameters around the electroweak scale ($\alpha_{\tilde t} M_C^2 \approx$ 
$\mbox{TeV}^2$) even when the compactification scale is as high as the GUT 
scale ($M_C \approx 10^{16}$ GeV).}:
\be
\mkstop^2[h_{u}] = \left[\left(k + \frac{m_t[h_u]}{M_C}\right)^2 +
\frac{\alpha_{\tilde t}}{\pi^2} + Z_k^2[h_u,\alpha_{\tilde t}] \right] M_C^2
 \left[ \vrule width 0pt height 18pt 1+ {\cal O}(\alpha_{\tilde t}^{2}) 
  \right] \qquad (k= -\infty,\dots,\infty)
 \label{mkstop_minimal}
\ee
where we have used the field-dependent top mass $m_{t}[h_{u}]$ from 
Eq.(\ref{eq:phystop}), and the function $Z_k^2[h_u,\alpha_{\tilde t}]$
is given by:
\be
 \begin{array}{rl}
  Z_{k\neq0}^{2}[h_u,\alpha_{\tilde t}] = &
   \left\{
    \begin{array}{lcl}
     \alpha_{\tilde t}/{\pi^2}  & \hspace*{5mm} & \mbox{if} <h_u> = 0 \\
      0			&	 &   \mbox{otherwise}
    \end{array}
   \right. \\
  Z_{k=0}^{2}[h_u,\alpha_{\tilde t}] = & 0 \\ 
\end{array}  \label{zfunction}
\ee
In the limit that electroweak symmetry remains unbroken
(i.e. $<h_{u}> = 0$), we find that only the even-parity bulk stop KK-modes
acquire soft masses from the non-renormalizable operators of
Eq.(\ref{nonren}), while the odd parity stops have the usual KK masses
($m_{\tilde{t},k}^{odd} = k M_{C}$).  Therefore the KK-summation only
runs over the positive tower ($0<k<\infty$).  Following EWSB ($<h_{u}>
\neq 0$), the Yukawa interaction of Eq.(\ref{yukawa}) induces mixing
between different parity stop KK-modes, and the KK-summation recovers
the full infinite tower.
Comparing Eqs.(\ref{eq:topmass}) and (\ref{mkstop_minimal}),
we can see that the form of the $k=0$ mass eigenvalues resemble the 
MSSM as expected~\footnote{For simplicity, we have neglected the trilinear 
mixing term $A_t$ between the stop left and right fields.} where the stop 
mass squared is given by the top mass and SUSY breaking mass added in 
quadrature.  However, in this limit the compactification scale 
$M_{C}$ can be at very high energies, and we recover the MSSM 
phenomenology at low energy.

Therefore, we consider maximal mixing in the stop sector 
($\alpha_{\tilde t} \gg 1$) which is equivalent to a large extra dimension.
In this case, the solution of Eq.~(\ref{x}) is given by
\be
 \mkstop^2 = \left(k + \frac{1}{2} \right)^2 M_C^2 
  \left[ \vrule width 0pt height 16pt
   1 + {\cal O} \left( \alpha_{\tilde t}^{-1} \right) \right] \qquad\quad 
    (k= -\infty, \dots, \infty)   \label{eq:stopmass} 
\ee  
Note that to leading order the stop mass eigenvalues are independent of
the Higgs background field ($h_{u}$) even when the electroweak
symmetry is broken. This can be attributed
to the arbitrarily large mixing term on the source brane that makes the
Yukawa brane become ``transparent'' which washes out any field
dependence~\cite{dkr}.  The Yukawa interaction induces mixing between 
different parity stop fields on the Yukawa brane such that both odd and 
even-parity stop KK-modes acquire the same mass from Eq.(\ref{eq:stopmass}).  
We find that the compactification scale should be $M_C\approx {\mathcal O} 
({\mathrm TeV})$
to provide soft terms at the electroweak scale~\footnote{This maximal mixing 
eigenvalue has a very weak dependence on the precise magnitude of the 
$\alpha_{\tilde t}$-mixing and so the $\alpha_{\tilde t}^{-1}$ corrections in
Eq.(\ref{eq:stopmass}) can be neglected.}.

We use dimensional regularization and zeta-function regularization 
techniques~\cite{kubyshin} to evaluate the top/stop contributions to the 
1-loop effective potential:
\begin{eqnarray}
 V_{1-loop} = V_{top} \, [h_{u}] + V_{stop}
  &=& \frac{1}{2} Tr \sum_{k=-\infty}^{\infty}
   \int \!\! \frac{d^{4}p}{ \left( 2 \pi \right)^{4}}  
    \ln \left[ \frac{ p^{2} + m_{\tilde{t},k}^{2}}
     {p^{2} + m_{t,k}^{2} \, [h_u]} \right]    \label{eq:v1loop} \\
  &=& \frac{9 M_{C}^{4}}{16 \pi^{6}} \sum_{n=1}^{\infty} 
   \frac{\cos \left[ 2 \pi \, n \, m_t[h_{u}]/M_C \right]}{n^{5}} \,\,
    + \,\, \frac{135 \, \zeta[5]}{256 \pi^{6}} M_{C}^{4}  \hspace*{1cm}
     \nonumber
\end{eqnarray}
where we have used Eqs.(\ref{eq:topmass}),(\ref{eq:phystop}) and 
(\ref{eq:stopmass}), and the trace is over all degrees of freedom.  The top 
and stop contributions are found to be separately finite (and free of 
ultraviolet divergences).  However, 
we find the same result by using other regularization techniques 
(i.e. KK-regularization~\cite{delgado,barbieri,arkani,quiros}), where the 
UV divergences vanish due to supersymmetric cancellations between the
top and stop contributions.   Note that only the top contribution is field 
dependent, so the constant stop contribution can be absorbed into the
cosmological constant and dropped from the subsequent analysis~\footnote{We 
show that the field-dependence in the stop sector vanishes using a 
diagrammatic approach in Appendix \ref{app:a}.}.

\subsection{SUSY Breaking Higgs Parameters}  \label{sec:higgs}

We have seen in section \ref{sec:kkspec} that we are led to maximal mixing 
($\alpha_{\tilde{t}} \gg 1$) in the stop sector.  We also know that for
EWSB via top/stop radiative corrections, we require a 
negative mass-squared to trigger spontaneous symmetry breaking.  This is 
harder to achieve when 
$\alpha_{\tilde{t}} \approx \alpha_{H}$ since this leads to tree-level and 
(negative) 1-loop contributions of comparable magnitude.  Therefore, we 
conclude that the KK-modes in the Higgs sector must be minimally mixed 
(i.e. $\alpha_{H} \ll 1$).

Using Eq.(\ref{alphas}), we have two options, 
either (i) the couplings in the higher-dimensional operators are hierarchical 
($c_{\tilde{t}} \gg c_{H}$), or else (ii) there is a non-univerality in the 
hidden sector where different SUSY breaking fields couple to the stop and 
Higgs sectors ($F_{S , H} \neq F_{S , \tilde{t}}$).  In this paper, we will 
assume that all hidden sector couplings are $c \sim {\mathcal O}(1)$, and
instead have non-universal F-terms ($F_{S , H} \ll F_{S , \tilde{t}}$).

For minimal mixing in the Higgs sector, the KK mass matrix is dominated by the
diagonal terms and we can decouple the non-zero KK excitations. We will
impose the EWSB conditions on the lightest ($k=0$) KK-modes where the soft 
masses are taken directly from Eq.(\ref{nonren}):
\begin{eqnarray}
 m_{H_{u}}^{2} = m_{H_{d}}^{2} = m_{soft}^{2}
  = \frac{c_{H} F_{S,H}^{2}}{M_{\ast}^{3} \, \pi} M_C 
\hspace*{5mm},\hspace*{5mm}
 B\mu = \frac{c_{B\mu} F_{S,H}^{2}}{M_{\ast}^{3} \, \pi } M_C 
\hspace*{5mm},\hspace*{5mm}
 \mu = \frac{c_{\mu} F_{S,H}}{M_{\ast}^{2} \, \pi } M_C  
    \label{softhiggs} 
\end{eqnarray}
and we have assumed that $c_{H_{u}}=c_{H_{d}}=c_{H}$ for universal soft Higgs 
masses. 

\subsection{Reliability and Perturbativity}  \label{sec:pert}

We have not yet imposed any constraint on the relationship between the 
compactification scale $M_C$ and the cutoff scale $M_{\ast}$.  The 
requirement of perturbativity (where our perturbative analysis is valid) 
allows us to find an upper bound on the ratio $(M_*/M_C)$.  It is well known
that in 5D (extra-dimensional) theories that the gauge and Yukawa couplings 
exhibit power law running behaviour~\cite{dudas}. The beta functions of these
couplings depend on powers of the renormalization scale $\mu$ due to the 
inclusion of the KK-modes that makes the physics highly sensitive to the
renormalization scale.  This implies that gauge coupling unification and the 
emergence of the Landau pole in the Yukawa couplings are accelerated with 
respect to the (logarithmically-running) 4D theory.  The top Yukawa coupling 
is found to become singular at energies close to the compactification scale 
$M_C$.  In our model, the presence of the third family in the 5D bulk makes 
the Yukawa coupling beta function ($\beta_{y_{t}}$) depend quadratically on 
the ratio between the renormalization scale $\mu$ and the compactification 
scale $M_{C}$
\begin{eqnarray}
 \beta_{y_{t}} \sim y_{t}^{3} \left( \frac{\mu}{M_C} \right)^{2} + \ldots
   \label{betayt}
\end{eqnarray}
Note that the dependence on $(\mu/M_C)$ is stronger than the corresponding 
beta function for the gauge coupling ($\beta_{y_{t}}$) which is only 
linearly-dependent
\begin{eqnarray}
 \beta_{g} \sim g^{3} \left( \frac{\mu}{M_C} \right) + \ldots
   \label{betag}
\end{eqnarray}

Suppose $\mu_{NP}$ is the scale where the top 
Yukawa coupling becomes non-perturbative, which we numerically found to be 
$\mu_{NP} \approx 5 M_C$ ~\cite{quiros}.  We can maintain a (reliable) 
perturbative regime by imposing the following constraint on the cutoff scale
of our theory $M_{\ast}$
\begin{eqnarray}
 M_{\ast} \simlt \mu_{NP} \hspace*{1cm} \longrightarrow \hspace*{1cm}
  M_{\ast} \simlt 5 M_C   \label{pert}
\end{eqnarray}
Similarly it appears ``unnatural'' to have the cutoff of our theory $M_{\ast}$
below the compactification scale $M_{C}$.  Hence, we find that perturbativity
and naturalness severely constrain $M_{\ast}$ to the range:
\begin{eqnarray}
  M_{C} \simlt M_{\ast} \simlt 5 M_{C}  \label{window}
\end{eqnarray}

We need to check that these constraints are consistent with maximal mixing in 
the stop sector ($\alpha_{\tilde t} \gg 1$). However, we find that there is no
inconsistency since our earlier work~\cite{dkr} showed numerically that 
maximal mixing only requires $\alpha_{\tilde t} \simgt 12$ since 
solutions of Eq.(\ref{x}) tend towards an asymptotic value with the stop mass 
eigenvalues given by Eq.(\ref{eq:stopmass}).  This is achieved when
$F_{S,\tilde{t}} \simgt (0.9/c_{\tilde t}) M_*^2 \approx M_{\ast}^{2}$ for 
$c_{\tilde t} \approx 1$. 
  
\section{Higgs Mass Spectrum}   \label{sec:2hdm}

In this section we calculate the mass eigenvalues in the Higgs sector, where
the light and heavy CP-even higgs mass eigenstates ($h^{0} , H^{0}$) are 
linear combinations of the real fields $h_{u}$ and $h_{d}$.  We can use the 
standard MSSM relations to find the masses of the charged ($H^{\pm}$) and 
CP-odd ($A^{0}$) Higgs fields.
The tree-level potential in terms of the neutral components of the
Higgs doublets ($H_{u}^{0} , H_{d}^{0}$) is:
\begin{eqnarray}
 V_{tree}=m_{1}^{2} \left| H_{d}^{0} \right|^{2} +
  m_{2}^{2} \left| H_{u}^{0} \right|^{2} - B \mu \left( H_{u}^{0}
  H_{d}^{0} + h.c. \right) + \frac{m_{Z^{0}}^{2}}{2v^{2}}
  \left[ \left| H_{u}^{0} \right|^{2} - \left| H_{d}^{0} \right|^{2}
  \right]^{2}   \label{eq:vtree}
\end{eqnarray}
where we are free to define $B\mu$ as real and positive by absorbing
any phase into $H_{u}^{0}$ and $H_{d}^{0}$; and we have traded the 
$U(1)_{Y}$ and $SU(2)_{L}$ gauge couplings ($g',g$) for the physical 
$Z^{0}$ mass and the VEV $v=246$ GeV.
The soft parameters $m_{1}^{2}=\left|\mu\right|^{2}+m_{H_{d}}^{2}$ ,
$m_{2}^{2}=\left|\mu\right|^{2}+m_{H_{u}}^{2}$ and $B\mu$ are given in
Eq.(\ref{softhiggs}).

\begin{figure}[h!]
\begin{center}
\psfrag{Mc}{{\small $M_{C}$ \, (GeV)}}
\psfrag{Msoft}{{\small \hspace*{-8mm} $m_{soft}$ (GeV)}}
\psfrag{Tan}{}
\psfrag{=}{}
\psfrag{Beta}{}
\psfrag{1.5}{{\large \hspace*{-1.5cm} $\tan\beta = 1.5$}}
\psfrag{20}{{\large \hspace*{-3cm} $\tan\beta = 20$}}
\scalebox{1.2}{
{\mbox{\epsfig{file=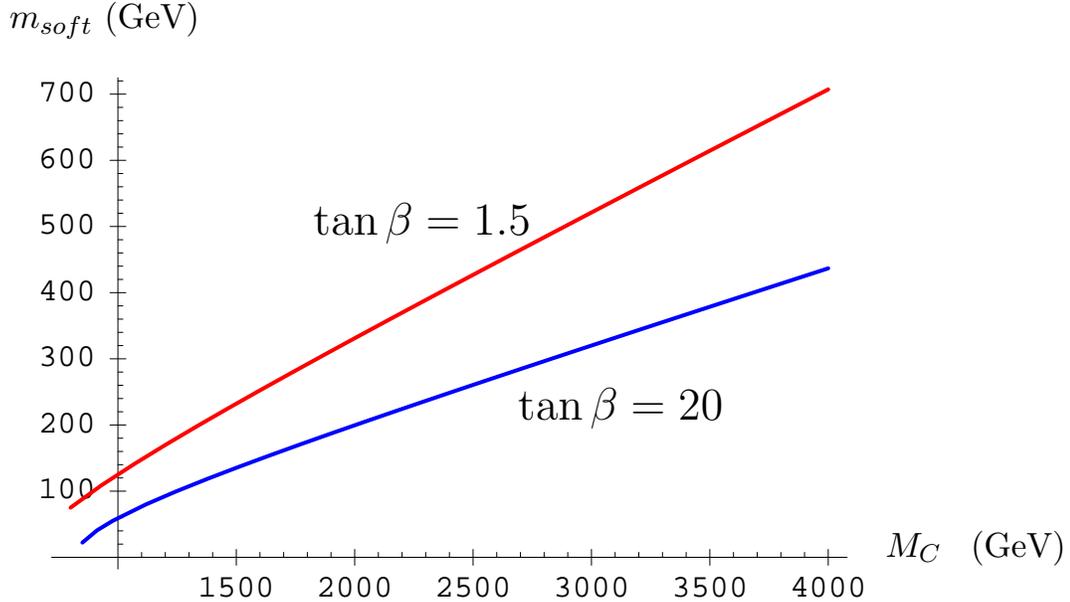}}}}
\caption{{\small The universal Higgs soft mass 
 $m_{soft}^{2}=\left| \mu \right|^{2} + m_{H_{u,d}}^{2}$ against the 
compactification scale $M_{C}$ for $\tan\beta = 1.5$ and $20$.}}
  \label{fig:msplot}
\end{center}
\end{figure}

Combining Eqs.
(\ref{eq:higgsdef},\ref{eq:v1loop},\ref{eq:vtree}), we find an
expression for the total 1-loop effective potential 
in terms of the real Higgs fields $h_{u} , h_{d}$.  Notice that we have 
dropped the 1-loop stop contribution since it is independent of the Higgs 
fields and can be absorbed into the cosmological constant.
\begin{eqnarray}
 V =  \frac{m_{soft}^{2}}{2} \left( h_{u}^{2} + h_{d}^{2} \right) 
  - B \mu \, h_{u} h_{d} + \frac{m_{Z^{0}}^{2}}{8 v^{2}}
  \left[ h_{u}^{2} - h_{d}^{2} \right]^{2}  
 + \frac{9 M_{C}^{4}}{16 \pi^{6}}
  \sum_{n=1}^{\infty} \frac{\cos \left[ 2 n \tan^{-1}
  \left( {\displaystyle \frac{y_{t} \, \pi \,  h_{u}}{M_{C} \, \sqrt{2}} }
   \right) \right]}{n^{5}}   \label{eq:vfull}
\end{eqnarray}
and we have assumed that the Higgs doublets acquire universal soft masses 
$m_{1}^{2}=m_{2}^{2}=m_{soft}^{2}$ which we regard as input parameters along 
with $B\mu$.
Applying the EWSB conditions at the usual minimum 
$\left< h_{u} \right> = v \sin\beta \, , \, \left< h_{d} \right> = v
\cos\beta$ \, ($v= 246$ GeV)
\begin{eqnarray}
 \left. \frac{\partial V}{\partial h_{u}} \right|_{\left< h_{u} \right>, 
\left< h_{d} \right>} = 
  \left. \frac{\partial V}{\partial h_{d}} \right|_{\left< h_{u} \right>, 
\left< h_{d} \right>} = 0
  \label{eq:min1}
\end{eqnarray}
allows us to eliminate $B\mu$ and $m_{Z^{0}}^{2}$ in terms of the
other parameters.  By imposing the
correct observable $Z^{0}$ mass ($m_{Z^{0}} = 91.2$ GeV), we can also
eliminate $m_{soft}$ such that the
compactification scale $M_{C}$ and $\tan\beta$ (or equivalently
$m_{A^{0}}$) can be regarded as the two free parameters.  In Figure
\ref{fig:msplot}, we plot $m_{soft}$ as a function of the compactification 
scale $M_{C}$ for two different values of $\tan\beta$. 
%

We construct the CP-even mass matrix from the second derivatives of
the effective potential at the minimum, which can be diagonalized to
find the mass eigenvalues ($m_{h^{0}} , m_{H^{0}}$):
\begin{eqnarray}
 {\mathcal M}^{2}_{even} = 
  \left( 
   \begin{array}{cc}
    {\displaystyle \frac{\partial^{2} V}{\partial h_{u}^{2}} }
  & {\displaystyle \frac{\partial^{2} V}{\partial h_{u} \partial h_{d}} }\\
%
    {\displaystyle  \frac{\partial^{2} V}{\partial h_{d} \partial h_{u}}} 
  & {\displaystyle \frac{\partial^{2} V}{\partial h_{d}^{2}} }
   \end{array}
  \right)   \label{eq:massmatrix}
\end{eqnarray}
Notice that these CP-even eigenvalues include the 1-loop effects, but only
$m_{2}^{2}$ is 1-loop improved since we are neglecting bottom-sbottom loops
for $\tan\beta \le 20$:
\begin{eqnarray}
 m_{2,imp}^{2} = m_{2}^{2} + \frac{1}{2} \left. 
  \frac{\partial^{2} V_{top}}{\partial h_{u}^{2}} 
   \right|_{\left< h_{u} \right>, \left< h_{d} \right>} 
    \label{eq:improved}
\end{eqnarray} 
However $B\mu$ (and $m_{1}^{2}$) are not 1-loop improved, so we can use the
standard 4D tree-level MSSM expressions to find the CP-odd ($A^{0}$) and
charged Higgs ($H^{\pm}$) masses.
\begin{eqnarray}
 m_{A^{0}}^{2} &=& \frac{2 B \mu}{\sin2\beta} \label{eq:ao} \\
 m_{H^{\pm}}^{2} &=& m_{A^{0}}^{2} + m_{W}^{2} \label{eq:hpm}
\end{eqnarray}
We can solve for $m_{Z^{0}}^{2}$ as a function of the free
parameters ($M_{C} , \tan\beta$), and use the standard definition of
fine-tuning $\Delta$ ~\cite{finetune,finetune2} (but neglecting the 
variation of $\tan\beta$ with respect to changes in $M_{C}$) to find an upper 
limit on the compactification scale $M_{C}$. 
\begin{eqnarray}
 \Delta = \left| \frac{M_{C}}{m_{Z^{0}}^{2}} \, 
   \frac{\partial m_{Z^{0}}^{2}}{\partial M_{C}} \right|  \label{eq:delta}
\end{eqnarray}
%
\begin{figure}[h!]
\begin{center}
\psfrag{Delta}{{\small \hspace*{0.5cm} $\Delta$}}
\psfrag{Tan}{{\small \hspace*{3mm} $\tan\beta$}}
\psfrag{Beta}{}
\psfrag{Mc}{{\small \hspace*{-2cm} $M_{C}$ (GeV)}}
\scalebox{1.3}{
{\mbox{\epsfig{file=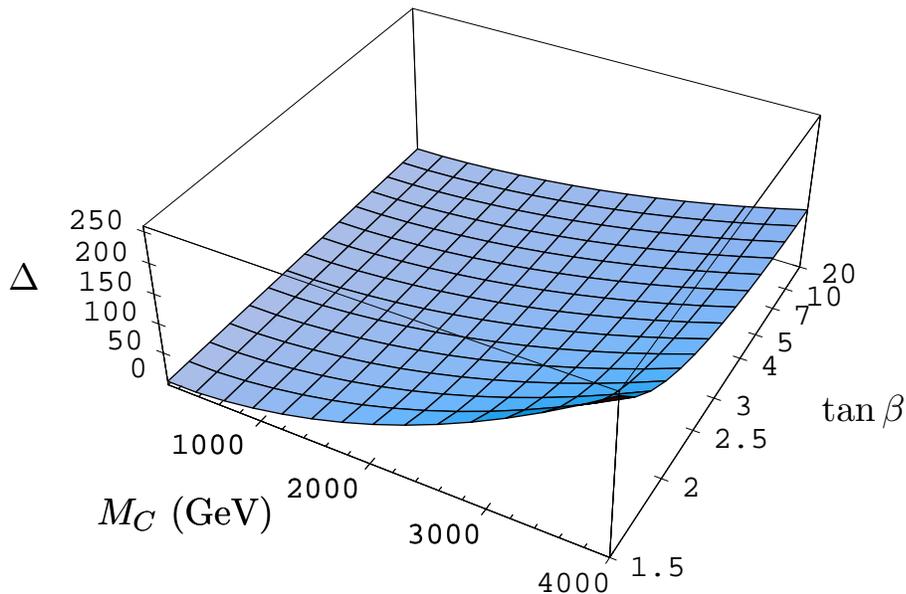}}}}
 \vskip-5mm
\caption{{\small The fine-tuning parameter $\Delta$ as a function of
$\tan\beta$ and the compactification scale $M_{C}$ in GeV.  $\Delta$ is shown
to have a very weak dependence on $\tan\beta$ in the range $2.5 \simlt
\tan\beta \simlt 20$ for which we can neglect bottom-sbottom effects.  
However, the fine-tuning becomes singular as $\tan\beta \rightarrow 1$.}}
  \label{fig:delta}
\end{center}
\end{figure}

\noindent Motivated by the fine-tuning as shown in Figure \ref{fig:delta},
we will investigate the parameter space $M_{C} \le 4$ TeV and 
$1.5 \le \tan\beta \le 20$
where the fine-tuning $\Delta \sim {\mathcal O}(10^{2})$.
\begin{figure*}[h!]
\begin{center}
\psfrag{pm}{$H^{\pm}$}
\psfrag{Higgs}{{\small Higgs Mass (GeV)}}
\psfrag{Mass}{}
\psfrag{Tan}{}
\psfrag{=}{}
\psfrag{Beta}{}
\psfrag{1.5}{{\large \hspace*{-3cm} $\tan\beta = 1.5$}}
\psfrag{20}{{\large \hspace*{-2cm} $\tan\beta = 20$}}
\psfrag{mA}{{\small $m_{A^{0}}$ (GeV)}}
\scalebox{1.1}{
\epsfig{file=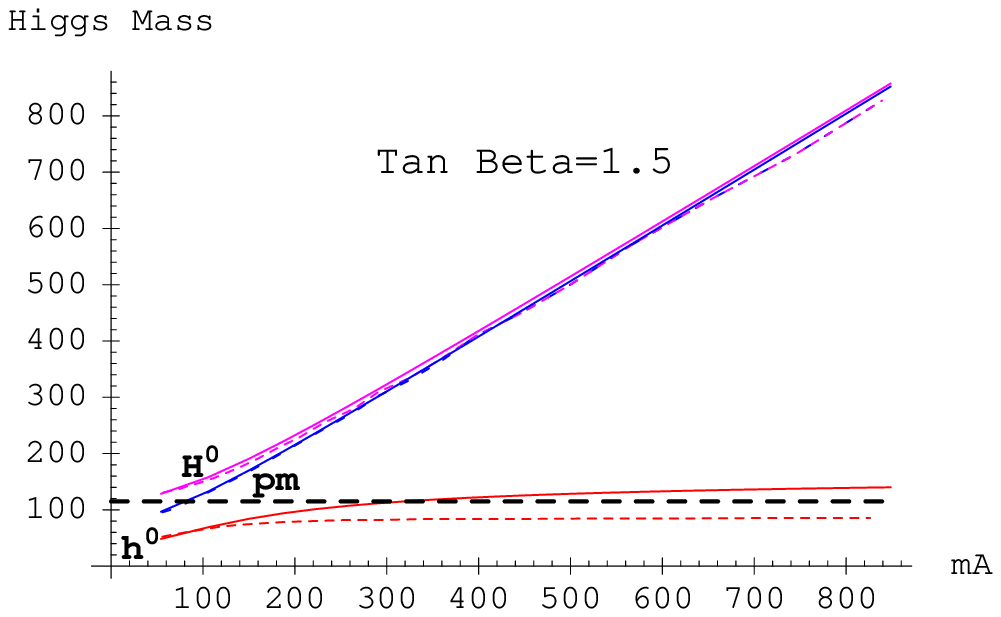}}\\
\scalebox{1.1}{\epsfig{file=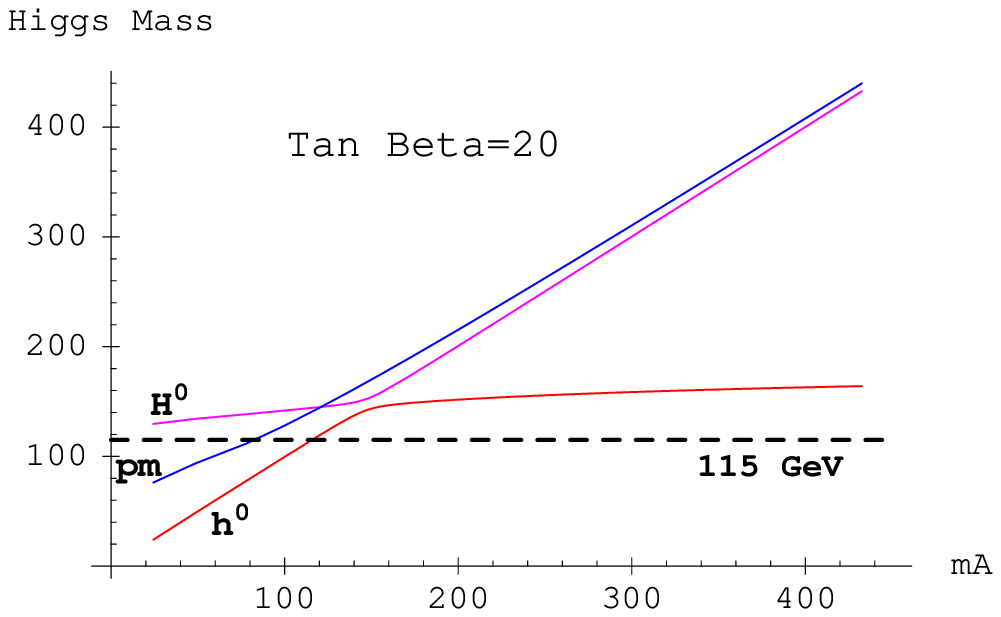}}
\caption{{\small The Higgs masses ($m_{h^{0}} ,
m_{H^{\pm}} , m_{H^{0}}$) against $m_{A^{0}}$ for $\tan\beta = 1.5$ (upper 
panel) and $\tan\beta = 20$ (lower panel).  We have run $1 \le M_{C} \le 4$ 
TeV parametrically along 
each curve.  The LEP candidate at 115 GeV ({\it bold dotted-line}) ~\cite{LEP}
is shown for comparison with $m_{h^{0}}$.
For $\tan\beta=1.5$ (upper panel), we also compare the MSSM results taken 
from \cite{pdg} ({\it dotted-lines}) against our model ({\it bold-lines}).}}
  \label{fig:higgs}
\end{center}
\end{figure*}

In Figure \ref{fig:higgs}, we plot the
eigenvalues ($m_{h^{0}} , m_{H^{0}}$) of the CP-even mass matrix 
and the charged Higgs mass ($m_{H^{\pm}}$)
against the CP-odd mass ($m_{A^{0}}$) for two fixed values of
$\tan\beta=1.5$ and $20$.  We run the 
value of the compactification scale parametrically along each curve, 
$1 \le M_{C} \le 4$ TeV.  We also include the MSSM predictions for 
$\tan\beta=1.5$ taken from \cite{pdg} for comparison. 
Notice that unlike the
MSSM predictions from \cite{pdg}, our model is {\it not} excluded by the LEP
signal for $\tan\beta=1.5$.  There are additional experimental
lower limits for the other Higgs masses
\begin{eqnarray}
 m_{A^{0}} \ge 92 \, {\mathrm GeV \hspace*{1cm} and} \hspace*{1cm} 
   m_{H^{\pm}} \ge 69 \, {\mathrm GeV}   \label{expt}
\end{eqnarray}
but these provide a much weaker constraint on our model.
\begin{figure}[h!]
\begin{center}
\psfrag{Mc}{{\small \hspace*{-6mm} $M_{C}$ (GeV)}}
\psfrag{Tan}{\vspace*{10mm}{\small \hspace*{6mm} $\tan\beta$}}
\psfrag{Beta}{}
\scalebox{1.2}{
{\mbox{\epsfig{file=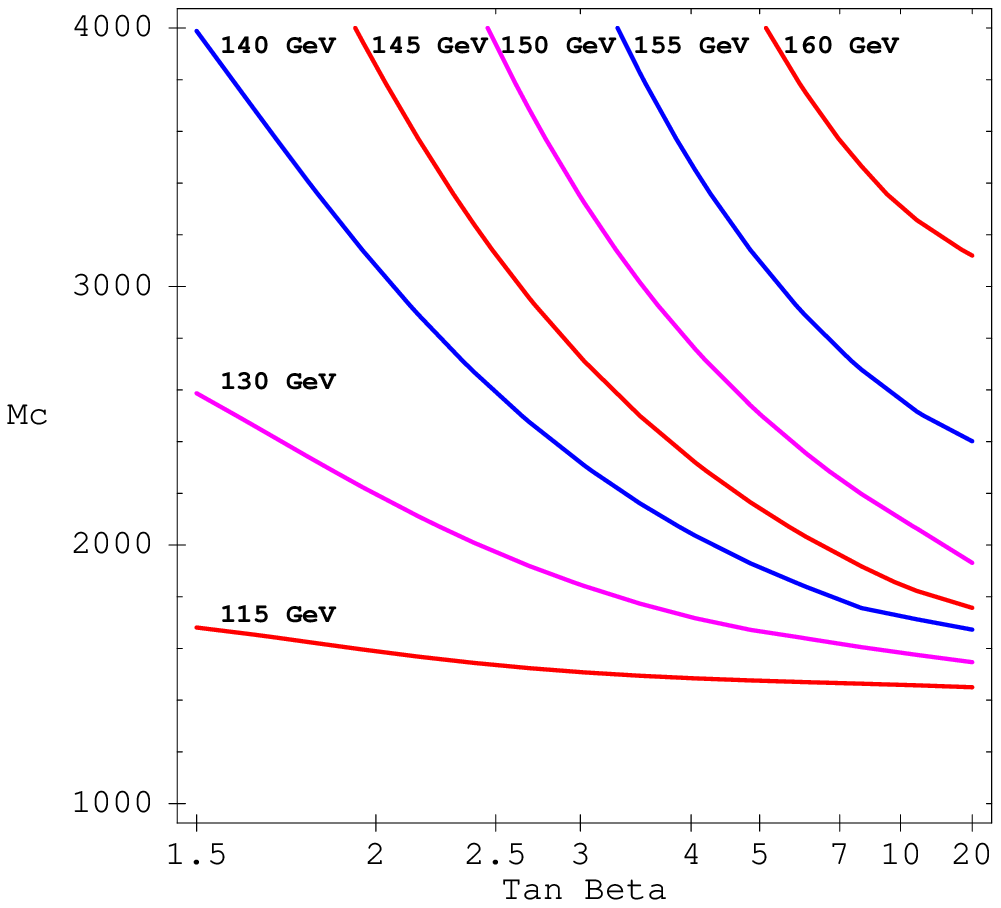}}}}
\vskip-5mm
\caption{{\small Mass contour plot for the lightest Higgs mass
$m_{h^{0}}$ as a function of the compactification scale $M_{C}$ (GeV) and
$\tan\beta$ in the absence of bottom sector effects (i.e. $\tan\beta
\le 20$).  The LEP candidate at 115 GeV ~\cite{LEP} is easily accommodated 
over the range of $\tan\beta$ with a
compactification scale $M_{C} \approx 1.5 - 1.7$ TeV.  At large $\tan\beta
\rightarrow 20$ and $M_{C} \rightarrow 4$ TeV, the lightest Higgs mass
can be as large as $m_{h^{0}} \sim 164$ GeV.}}
  \label{fig:mhcomp}
\end{center}
\end{figure}
 
In Figure \ref{fig:mhcomp}, we plot the lightest Higgs mass ($m_{h^{0}}$)
as a function of the compactification scale $M_{C}$ and $\tan\beta$.
The LEP data excludes the parameter space below the first contour at
$m_{h^{0}} = 115$ GeV ~\cite{LEP} which corresponds to a compactification 
scale $M_{C}  \approx 1.5 - 1.7$ TeV over the whole range of $\tan\beta$.
Combining Figures \ref{fig:delta} and \ref{fig:mhcomp} we find an allowed
window for the compactification scale
\begin{eqnarray}
  1.5 \, {\mathrm TeV} \simlt M_{C} \simlt 4 \, {\mathrm TeV}
\end{eqnarray}

Notice that our model can easily accommodate the conventional 4D MSSM upper
bound on the lightest Higgs boson mass $m_{h^{0}} \sim 130$ GeV, and can be
pushed as high as $m_{h^{0}} \sim 160$ GeV with $M_{C} \approx 4$ TeV and
$5 \simlt \tan\beta \simlt 20$.  Remember that including additional matter
(e.g. gauge singlets in the NMSSM) can also raise the upper bound on
the lightest Higgs mass, but our ``minimal'' extension of the MSSM achieves 
higher mass bounds without adding extra matter content.

We can use Figure \ref{fig:msplot} to find limits on the 
$\mu$-parameter from the universal soft Higgs mass $m_{soft}$ which is
constrained when we impose EWSB at the electroweak minimum.
Using Eq.(\ref{softhiggs}), we find the ratio
between $m_{H_{u}}^{2} = m_{H_{d}}^{2} = m_{H}^{2}$ and 
$\left| \mu \right|^{2}$ in terms of the compactification and cutoff scales:
\begin{eqnarray}
 \frac{m_{H}^{2}}{\left| \mu \right|^{2}} \sim \frac{M_{\ast} \pi}{M_{C}} 
   \label{ratio}
\end{eqnarray}
where we have assumed that $c_{H} , c_{\mu} \sim 1$.

Recall from section \ref{sec:higgs} that we assume
non-universality in the SUSY breaking sector 
($F_{S,H} \ll F_{S,\tilde{t}}$)
to obtain maximal (minimal) mixing in the stop (Higgs) sectors for viable 
radiative EWSB without hierarchical couplings $c_{\tilde{t}} \sim c_{H}$.
We can use the perturbativity and naturalness constraints of
Eq.(\ref{window}) to find limits on the ratio between the soft Higgs
mass and the $\mu$-parameter from Eq.(\ref{ratio})
\begin{eqnarray}
 \pi \simlt \frac{m_{H}^{2}}{\left| \mu \right|^{2}} \simlt 5\pi   
  \label{window2}
\end{eqnarray}
and therefore limits on $m_{soft}$ in terms of $\mu$:
\begin{eqnarray}
 m_{soft}^{2} = m_{H}^{2} + \left| \mu \right|^{2}
 \hspace*{1cm} \longrightarrow
  \hspace*{1cm} (\pi + 1) \, \left| \mu \right|^{2} \simlt m_{soft}^{2}
   \simlt (5\pi + 1) \, \left| \mu \right|^{2}
\end{eqnarray}

\noindent The constraints on 
the $\mu$-parameter for different compactification scales and values of 
$\tan\beta$ are shown in Table \ref{tab:mulimits}:
\begin{table}[h]
 \begin{center}
  \begin{tabular}{|c||c|c|} \hline
  & $M_{C}=1.5 \, {\mathrm TeV}$ & $M_{C}=4 \, {\mathrm TeV}$ \\ \hline\hline 
     \vrule width 0pt height 13pt
  $\tan\beta = 1.5$ & $114 \simgt \left| \mu \right| 
   \simgt 57$ & $347 \simgt \left| \mu \right| \simgt 173$  \\ \hline
     \vrule width 0pt height 13pt
  $\tan\beta = 20$ & $66 \simgt \left| \mu \right| \simgt 33$ 
   & $215 \simgt \left| \mu \right| \simgt 107$ 
    \\ \hline
 \end{tabular}
  \caption {{\small Upper and lower limits on the size of the $\mu$-parameter
for two different values of $\tan\beta = 1.5$ and 20, and 
compactification scales $M_{C} = 1.5$ and 4 TeV.}}
  \label{tab:mulimits} 
 \end{center}
\end{table}

\noindent Therefore, the magnitude of the $\mu$-parameter is constrained to 
the range $33 \simlt \left|\mu\right| \simlt 347$ GeV. 

\section{Conclusions and Discussion}  \label{sec:conc}

In conclusion we have considered the Higgs sector of an ${\mathcal
N}=1$ supersymmetric 5D theory compactified on an orbifold
$S^{1}/Z_{2}$ where the compactification scale 
$M_{C} \sim {\mathcal O} \left( TeV \right)$.
Orbifolding leads to fixed points at either end of the extra dimension
(y) where 4D branes can be localized.  
Supersymmetry is broken by the F-term VEV of a gauge singlet on the
brane at $y=\pi R$ that is spatially separated from another Yukawa
brane at $y=0$ where the first two MSSM families and Yukawa couplings
are localized.  Direct coupling between
the two sectors (and therefore soft squark masses) are suppressed by
the separation between the branes which alleviates the flavour-changing
neutral-current problem since the first and second family squark
masses are only generated through flavour-blind loops.  The third family, 
gauge sector and Higgs
fields live in the extra dimensional bulk with their ${\mathcal N}=2$
supersymmetric partners (which are required for consistency) and therefore
receive unsuppressed soft masses due to their direct coupling to the
SUSY breaking sector.

We assume a non-universality in the SUSY breaking sector, where different 
gauge singlets couple separately to the stop (Higgs) fields and the 
associated F-term VEVs are hierarchical ($F_{S,H} \ll F_{S,\tilde{t}}$) which
induces maximal (minimal) mixing between different KK-modes.  The maximal 
mixing between stop modes requires that we use a matrix method to diagonalize 
the infinite mass matrix.  In contrast the Higgs KK-modes are
minimally mixed so that the mass matrix is dominated by the diagonal 
components and we can decouple the non-zero KK-modes from the
analysis.  We find that the soft Higgs parameters are generated by
non-renormalizable operators.
The presence of the third family in the bulk is particularly important
for their dominant 1-loop contribution to the Higgs effective 
potential.  The full tower of top and stop
Kaluza-Klein modes contribute to the potential and trigger
radiative electroweak symmetry breaking.  Following dimensional
regularization and zeta-function regularization techniques, we 
see that the 1-loop contributions to the effective potential are
separately finite and therefore insensitive to the high-energy cutoff 
$M_{\ast}$.  However in the maximal mixing limit we find that the top
contribution has a non-trivial dependence on the Higgs background fields,
and the stop only contributes to the cosmological constant. 

We minimize the 1-loop effective potential and impose the conditions for 
electroweak symmetry breaking to find the physical Higgs mass eigenvalues.
Requiring the correct physical $Z^{0}$-mass allows us to eliminate parameters
in terms of the compactification scale $M_{C}$ and $\tan\beta$ (or 
equivalently $m_{A^{0}}$).  We use fine-tuning arguments to constrain our
parameter space $M_{C} \simlt 4$ TeV, and we choose to study the region
$1.5 \simlt \tan\beta \simlt 20$ where bottom sector effects can be neglected.
We obtain physical Higgs mass eigenvalues for different values of $\tan\beta$
and find that the LEP signal~\cite{LEP} imposes a lower limit on the 
compactification scale $M_{C} \simgt 1.5$ TeV.  We also find that, 
unlike the MSSM, $\tan\beta=1.5$ is not excluded by experiment and our model
can accommodate the LEP signal over the full parameter space.  In fact the 
usual MSSM upper bound ($m_{h^{0}} \simlt 130$ GeV) and the extended NMSSM 
bound ($m_{h^{0}} \simlt 150$ GeV) can be trivially exceeded and 
raised to $m_{h^{0}} \simlt 164$ GeV for a compactification scale 
$M_{C} \sim 4$ TeV and $5 \simlt \tan\beta \simlt 20$.

Note that radiative electroweak symmetry breaking is viable over a large 
range $1.5 \simlt \tan\beta \simlt 20$ in comparison to an alternative model
~\cite{quiros} that is severely constrained to the smaller range
$35 \simlt \tan\beta \simlt 40$.  In fact, we show in Appendix 
\ref{app:b} that their diagrammatic analysis is incomplete because it 
neglects higher-order non-renormalizable operators that
inevitably appear in the expansion of the 1-loop effective potential.  
The requirement of perturbativity and naturalness in our model imposes a 
constraint on the relationship between the compactification scale $M_{C}$ 
and the cutoff $M_{\ast}$ scale ($M_{C} \simlt M_{\ast} \simlt 5 M_{C}$).  
We use this constraint in combination with the universal soft Higgs mass to 
deduce limits on the $\mu$-parameter, and we find that the magnitude
of $\mu$ is inside the range $33 \simlt \left|\mu\right| \simlt 347$ GeV.

Scherk-Schwarz (SS) boundary conditions have been used 
extensively in the literature to break SUSY in extra dimensional 
models~\cite{delgado,barbieri,arkani,quiros,masiero}.  Recently, the authors
of Ref.~\cite{riotto} demonstrated that the SS effects are always
present, and a vanishing SS parameter at tree-level will be generated at 
1-loop by bulk supergravity fields~\footnote{We thank Antonio Riotto for 
clarifying this point.}.  In the case that SUSY breaking is localized on a 
hidden sector brane, the SS breaking parameter $\omega_{SS}$ is no longer 
discrete, but may take a range of values 
that depend upon the relative strength of the SUSY breaking and the matter 
content of fields living in the extra 
dimension~\footnote{Ref.~\cite{riotto} investigate two scenarios with SUSY
breaking localized on hidden sector brane where the value of $\omega_{SS}$
depends on the relationship between $N_{H}$ and $N_{V}$, where 
$N_{H}$ ($N_{V}$) are the number of hypermultiplets (vector multiplets) in 
the bulk.}.  In our
model, the third family and Higgs hypermultiplets ($N_{H}=19$) live in the 
bulk with the MSSM vector multiplets ($N_{V}=12$), and the SS breaking 
parameter tends towards $\omega_{SS}=0$ ($\omega_{SS}=1/2$) for weak (strong)
SUSY breaking on the brane which is equivalent to minimal (maximal) mixing in
the stop sector.  

We anticipate that non-trivial SS contributions will modify our expressions 
for the stop KK-mode mass eigenvalues~\footnote{Remember that the top spectrum
remains unaffected by Scherk-Schwarz boundary conditions.} given in 
Eqs.(\ref{mkstop_minimal},\ref{eq:stopmass}).  However, we can see that the 
additional SS contributions will not alter our analysis of the Higgs sector. 
For minimal 
mixing the stop mass eigenvalues are given by Eq.(\ref{mkstop_minimal}), but 
we know that in this limit the SS parameter vanishes.  Also, for maximal
mixing $\omega_{SS} \rightarrow 1/2$ and Eq.(\ref{eq:stopmass}) becomes 
$m_{\tilde{t},k}^{2} = \left( k + 1/2 + \omega_{SS} \right)^{2} M_{C}^{2} 
+ \ldots$.  However, we have already seen that the stop mass eigenvalues are
independent of the background Higgs field in this maximal mixing limit and 
the stop sector contribution to the effective potential is constant.
Therefore, we find that Scherk-Schwarz effects play no r\^{o}le in our 
calculations and we are justified in ignoring them.

\begin{center}
{\bf \large Acknowledgements}
\end{center}
S.K., V.D.C. and D.R. would like to thank PPARC for a Senior Fellowship,
Research Associateship and a Studentship.  We would like to thank Antonio
Riotto for useful discussions and bringing Ref.~\cite{riotto} to our 
attention.  

\appendix
\section{Finite Higgs Mass from a Diagrammatic Approach}  
 \label{app:a}

\begin{figure}[h!]
 \begin{center} 
  \begin{picture}(450,70)(-265,-35)
  \DashLine(-265,10)(-229,10){3} \Vertex(-229,10){2}
  \Text(-255,20)[b]{$h_{u}$}
  \DashCArc(-200,10)(29,0,360){3}
  \Text(-200,51)[b]{$\tilde{t}_{L,k}^{mc} \, (\tilde{t}_{R,k}^{mc})$}
  \Text(-200,-23)[t]{$\tilde{t}_{R,l} \, (\tilde{t}_{L,l})$}
  \DashLine(-171,10)(-135,10){3} \Vertex(-171,10){2}
  \Text(-145,20)[b]{$h_{u}$}
  \Text(-255,-10)[b]{(a)}
  \DashLine(-105,10)(-69,10){3} \Vertex(-69,10){2}
  \Text(-95,20)[b]{$h_{u}$}
  \CArc(-40,10)(29,0,360)
  \Text(-40,51)[b]{$t_{L,k}$}
  \Text(-40,-23)[t]{$t_{R,l}$}
  \DashLine(-11,10)(25,10){3} \Vertex(-11,10){2}
  \Text(15,20)[b]{$h_{u}$}
  \Text(-95,-10)[b]{(b)}
  \DashLine(55,-19)(120,-19){3} 
  \Text(65,-9)[b]{$h_{u}$}
  \DashCArc(120,10)(29,0,360){3} \Vertex(120,-19){2}
  \Text(120,47)[b]{$\tilde{t}_{L,k} \, (\tilde{t}_{R,k})$}
  \DashLine(120,-19)(185,-19){3} \Text(175,-9)[b]{$h_{u}$}
  \Text(65,20)[b]{(c)}
 \end{picture}
\caption{One-loop diagrams contributing to $\delta m_h^2$ in 
Eq.(\ref{higgs0}).  Notice that the second and third diagrams (b,c) appear in
the MSSM while the requirement for ${\mathcal N}=2$ SUSY in the extra 
dimensional bulk leads to an additional diagram (a) involving CP-mirror 
partners.}   \label{feynman}
 \end{center}
\end{figure}
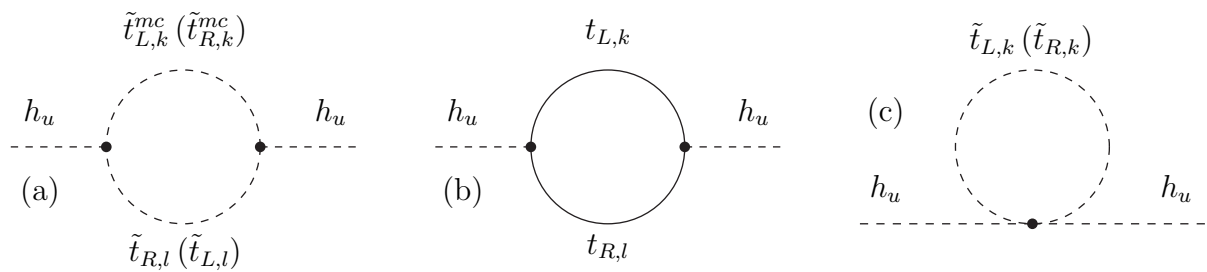

In section \ref{sec:kkspec} we have shown that in the limit the
 brane-localized SUSY 
breaking term associated with the stop sector is arbitrarily large
(maximal mixing), the 1-loop effective potential only receives 
field-dependence from the top sector KK masses while the stop sector 
contribution can be absorbed into the cosmological constant.  In this
appendix we will show that the finite part of the 1-loop Higgs scalar 
two-point function at zero external momenta, as shown in Figure~\ref{feynman},
only arises from the top KK-mode contributions. We illustrate this with a toy 
model that we have already discussed in earlier work~\cite{dkr}.
\vspace*{10mm}
 
In Ref.~\cite{dkr} we made the simplifying assumption that the tree level 
potential only receives D-term contributions by ignoring the Higgs soft 
parameters that are generated by non-renormalizable operators in 
Eq.(\ref{nonren}).  We made a further simplification by taking the limit
$\tan\beta \rightarrow \infty$ where the $H_{d}$ doublet can be decoupled. 
Therefore, the tree-level potential is only a function of the neutral 
component of the up-type Higgs $H_{u}^{0}$ which we take to be real $h_{u}$
from Eq.(\ref{eq:higgsdef})
\be
 V_{tree}[h_u] = \frac{g^2 + g'^2}{32} h_u^4
\label{tree}
\ee
where $g$ and $g'$ are the gauge couplings of $SU(2)_L$ and $U(1)_Y$
respectively.  We can see that at tree-level the potential has a minimum at 
$h_u = 0$.
However, the 1-loop contributions can trigger electroweak symmetry to be 
spontaneously broken. The 1-loop contribution is given in 
Eq.(\ref{eq:v1loop}). The full effective potential is then
\be
 V[h_u] = V_{tree}[h_u] + V_{1-loop}[h_u]   \label{full}
\ee
We find that a negative quadratic mass term is given by:
\be
 \delta m_h^2 = \left.\frac{d^2 V}{d h_u^2}\right|_{<h_u>=0} = - \frac{9\,
  \zeta(3)}{8\pi^4}\, y_t^2 M_C^2 \approx - \left(1.39\times 10^{-2}\right)\,
  y_t^2 M_C^2  \label{higgs0}
\ee
where $\zeta(3)\approx 1.202$. Notice that Eq.(\ref{higgs0}) is identical to
the result reported in Ref.~\cite{arkani}, where 
$\delta m_h^2$ was calculated using the diagrams shown in 
Figure~\ref{feynman}.  The mass eigenvalues of the even-parity stop KK-modes
($\tilde t_{R,l}$ and $\tilde t_{L,l}$) are given in Eq.~(\ref{eq:stopmass});
and the masses for the odd-parity (mirror) stops 
($\tilde t_{R,k}^{mc}$ and $\tilde t_{L,k}^{mc}$) and the tops
($t_{L,k}, t_{R,k}$) are $\sim k M_C$.  

It is possible to see that the finite part of $\delta m_h^2$
effectively arises from diagram (b) in Figure \ref{feynman} where
only the top KK-modes propagate in the loop. Evaluating diagrams (a)+(c) 
gives:
\bea
-i\delta m_h^2 \left[ \vrule width 0pt height 16pt 
 \mbox{(a)+(c)}\right] &=& 3y_t^2\sum_{k=0}^{\infty}\sum_{l=0}^
{\infty}\left(\eta_k\right)^2\int\!\! \frac{d^4 p}{(2\pi)^4} \frac{m^2_{\tilde t_{L,k}^{mc}}}{(p^2 - m^2_{\tilde t_{L,k}^{mc}})(p^2 - m^2_{\tilde t_{R,l}})}
+ (L\leftrightarrow R) \nonumber \\
&+& 3y_t^2\sum_{k=0}^{\infty}\sum_{l=0}^
{\infty}\left(\eta_k\right)^2\int\!\! \frac{d^4 p}{(2\pi)^4} \frac{1}{
(p^2 - m^2_{\tilde t_{R,l}})} + (L\leftrightarrow R)
\label{feo}
\eea 
where the factor of $3$ is the colour multiplicity and 
$\eta_k = (1/\sqrt{2})^{\delta_{k0}}$. Numerous papers in the 
literature~\cite{kubyshin,criticas} have demonstrated the validity of 
exchanging the sum and integral in Eq.(\ref{feo}) despite the controversy 
surrounding this so-called ``KK-regularization'' technique~\cite{nilles}.
Performing a Wick rotation to Euclidean 
momentum space $p_E$, and a change of variables $x= p_E/M_C$ gives
\bea
-i\delta m_h^2 \left[ \vrule width 0pt height 16pt
 \mbox{(a)+(c)} \right] &=& - i3y_t^2 M_C^2 \int\!\! \frac{d^4 x}
{(2\pi)^4} \, \sum_{k,l=0}^{\infty}\frac{(\eta_k)^2 x^2}{(x^2+(k+1/2)^2)(x^2+l^2)} \nonumber \\
&=& - i 3y_t^2 M_C^2\frac{\pi^2}{4} \int\!\! \frac{d^4 x}{(2\pi)^4} = 
  - \frac{3 \, i \, y_{t}^{2}}{128} \frac{\Lambda^{4}}{M_{C}^{2}} 
   \label{guapo}
\eea
where $\Lambda$ is the UV cutoff.
The quartically-divergent contribution in Eq.(\ref{guapo}) exactly cancels 
with the infinite part of diagram (b).  Therefore, $\delta m_{h}^{2}$ in
Eq.(\ref{higgs0}) only arises from diagram (b)~\footnote{Notice that this
is only true with maximal mixing in the stop sector.  For example, when
SUSY remains unbroken ($\alpha_{\tilde t}=0$), we obtain a finite 
contribution from diagrams (a)+(c) which cancels exactly with the 
contribution from diagram (b) such that $\delta m_h^2 = 0$.  Also note that 
when SUSY is (only) broken by Scherk-Schwarz boundary conditions,
the finite contribution to $\delta m_{h}^{2}$ arises from all three diagrams
in Figure~\ref{feynman} ~\cite{delgado,quiros}.}.

We will conclude this appendix by calculating the compactification scale and
Higgs mass in this toy model. We use the $\overline{{\mbox MS}}$ running top 
mass $m_{t,0} = 166$ GeV to
find the compactification scale by imposing the following minimization 
conditions around the minimum at $v=246$ GeV:
\begin{eqnarray}
 \left. \frac{ d V}{d h_{u}} \right|_{<h_{u}>=v} = 0 \hspace*{2cm}  
 \left. \frac{ d^{2} V}{d h_{u}^{2}} \right|_{<h_{u}>=v} 
   = m_{h}^{2}  \label{vmin}
\end{eqnarray}

\noindent Numerically we find:
\begin{equation}
 M_C \approx 830 \,\,{\mathrm GeV}
\label{Mc}
\end{equation} 
which is approximately $2.5$ times larger than the compactification scale 
calculated in
Ref.~\cite{barbieri}.  The second-derivative of the effective potential at
the minimum yields a lightest Higgs scalar mass~\footnote{Note that 
the published result in Ref. \cite{dkr} differs by a factor of $\sqrt{2}$.}
\begin{equation}
  m_{h} \approx 120 \,\, {\mathrm GeV}
\label{mh}
\end{equation}

This prediction for the lightest Higgs boson mass is possible because 
the effective potential just depends on the compactification scale $M_C$ 
which has been fixed at a specific value $M_{C} \approx 830$ GeV to obtain
the correct minimum of the effective potential.  However, the more general
two-Higgs doublet analysis in section \ref{sec:2hdm} depends on other 
soft parameters including $\tan\beta$, the universal soft Higgs mass 
$m_{soft}^{2} \, , \, B\mu$ and the supersymmetric mass $\mu$. 

\section{Truncation of the 1-loop Effective Potential}   \label{app:b}

In this appendix we discuss the problems that arise when truncating the 
expression of the full 1-loop effective potential in the
context of extra dimensions.  We will show that this truncation leads to a 
significant discrepancy compared to using the full 1-loop expression.  This
problem arises because all dimensionful parameters defined in the theory are 
of the same order $\sim M_{C}$.  In the effective field theory, it is 
impossible to distinguish between high and low energy scales, and all 
non-renormalizable operators are found to be as important as the 
renormalizable operators.   

Expanding the 1-loop effective potential~(\ref{eq:v1loop}) around the
origin $h_u = 0$ we obtain:
\be
 \tilde V_{1-loop}[h_u] = \frac{9 \, \zeta(5)}{16 \pi^6}M_C^4 + 
  \frac{1}{2}\delta m_h^2 \, h_u^2 + \frac{3 \, y_t^4}{16\pi^2} h_u^4 
   \left[\zeta(3) + \frac{25}{24} - \frac{1}{2}\log\left(\frac{\sqrt{2} \pi 
    y_t \, h_u}{M_C}\right)\right] + {\cal O}\left(\frac{h_u^6}{M_C^2}\right)
\label{expansion}
\ee
where we have used the expression for $\delta m_h^2$ from Eq.(\ref{higgs0})
and tildes denote truncated results. 
Notice that all non-renormalizable operators 
${\cal O}\left(h_u^6/M_C^2\right)$ vanish in the limit 
$M_C\rightarrow \infty$. 

However, for $M_C\approx {\mathcal O} ({\mathrm TeV})$, these higher-order 
operators will give an important contribution to the effective potential. 
Therefore, if the expansion is truncated at ${\mathcal O} (h_u^4)$ we can no 
longer regard the results as reliable. For example, if we
impose the minimization conditions of Eq.(\ref{vmin}) on the truncated 
potential from Eq.(\ref{expansion}) we find:
\be
\tilde M_C \approx 1 \,\,\mbox{TeV} \qquad\quad \tilde m_h \approx 170 \,\,\mbox{GeV}
\ee
Comparing with the exact results from Eqs.(\ref{Mc},\ref{mh}), we see 
that the truncated approximation from Eq.(\ref{expansion}) leads to an error 
of $(20-40)\%$. 

We can easily understand the origin of this discrepancy by 
noting that the only dimensionful parameter that appears in the effective 
potential is the compactification scale $M_C$.  Therefore, each  
operator in the expansion of Eq.(\ref{expansion}) gives comparable 
contributions.
Notice that the truncation of the series is equivalent to
integrating out the top KK-tower. From the effective field theory 
perspective, this truncation must be controlled by an expansion in 
$\gamma = (m^2/M^2)$, where $m$ is the low energy scale (for example the 
electroweak scale $M_{W}$) and $M$ is the high energy scale associated with 
the masses of the heavy particles $\sim M_{C}$.

In the case that $\gamma \ll 1$, the truncation of the series is reliable. 
However, in this model $m$ is generated at 
1-loop and is proportional to $M=M_{C}$, where $m^2 = \delta m_{h}^{2}$ from
Eq.(\ref{higgs0}) but without the loop factor~\footnote{Note that all of the 
loop factors in the expansion from Eq.~(\ref{expansion}) factorize out.}.
Therefore $m^2 \approx M^2 \rightarrow \gamma \approx 1$, and we conclude that
truncation of the effective potential expansion at finite order cannot be
reliable.  The same conclusion was observed in Ref.~\cite{barbieri}. 


\end{document}